\documentclass[letterpaper,twocolumn,10pt]{article}
\usepackage{usenix2019,epsfig,endnotes}
\usepackage{booktabs} 
\usepackage{algorithm}
\usepackage{algpseudocode}
\usepackage{mdframed}
\usepackage{array}
\usepackage{tabularx}
\usepackage{nicefrac}
\usepackage{balance}
\usepackage{subcaption}
\usepackage{threeparttable}
\usepackage{adjustbox}
\usepackage{amssymb}
\usepackage{pifont}
\usepackage{lipsum}
\usepackage{float}
\usepackage{comment}
\usepackage{xcolor}
\usepackage{amsmath}
\usepackage{amsthm}
\usepackage[hyphens]{url}
\usepackage{pifont}
\usepackage{hyperref}
\usepackage{xspace}
\usepackage{enumitem}

\begin{document}

\newcolumntype{L}[1]{>{\raggedright\arraybackslash}p{#1}}
\newcolumntype{C}[1]{>{\centering\arraybackslash}p{#1}}
\newcolumntype{R}[1]{>{\raggedleft\arraybackslash}p{#1}}
\newcommand{\cmark}{\ding{51}}%
\newcommand{\xmark}{\ding{55}}%

\theoremstyle{definition}
\newtheorem{recommendationt}{Recommendation}
\newcommand{\recommendation}[1]{\begin{mdframed}[backgroundcolor=gray!15]{\vspace{-0.07in}\begin{recommendationt}#1\end{recommendationt}\vspace{0.05in}}\end{mdframed}}

\newcommand{\ie}{{\it i.e.,}\xspace}
\newcommand{\eg}{{\it e.g.,}\xspace}

\newcommand{\absneed}[1]{{#1}}
\newcommand{\abstask}[1]{{#1}}
\newcommand{\absobject}[1]{{#1}}
\newcommand{\absfindings}[1]{{#1}}
\newcommand{\absconclusion}[1]{{#1}}

\newcommand{\attackName}{\textsc{UnderRadar}\xspace}

\newcommand\blfootnote[1]{%
  \begingroup
  \renewcommand\thefootnote{}\footnote{#1}%
  \addtocounter{footnote}{-1}%
  \endgroup
}

\newcommand{\paragraphbb}[1]{\vspace{0.05in}\noindent\textbf{{#1}}}
\newcommand{\parai}[1]{\vspace{0.05in}\noindent\textit{{#1}}}

\newcommand{\todo}[1]{\textcolor{red}{\textbf{TODO: {#1}}}}

\date{}

\title{\Large \bf Order P4-66\thanks{In the Star Wars universe, ``Order 66'' activated a secret program within the Republic's clone troopers, turning them against their former masters.} ~: Characterizing and mitigating surreptitious programmable~network device exploitation}

\author{
    {\rm Simon Kassing$^1$, Hussain Abbas$^1$, Laurent Vanbever$^1$, Ankit Singla$^1$}\\
    $^1$ETH Z\"urich
}

\maketitle

\thispagestyle{empty}

\subsection*{Abstract}
{\em Substantial efforts are invested in improving network security, but the threat landscape is rapidly evolving, particularly with the recent interest in programmable network hardware. We explore a new security threat, from an attacker who has gained control of such devices. While it should be obvious that such attackers can trivially cause substantial damage, the challenge and novelty are in doing so while preventing quick diagnosis by the operator. 

We find that compromised programmable devices can easily degrade networked applications by orders of magnitude, while evading diagnosis by even the most sophisticated network diagnosis methods in deployment. Two key observations yield this result: (a) targeting a small number of packets is often enough to cause disproportionate performance degradation; and (b) new programmable hardware is an effective enabler of careful, selective targeting of packets. Our results also point to recommendations for minimizing the damage from such attacks, ranging from known, easy to implement techniques like encryption and redundant requests, to more complex considerations that would potentially limit some intended uses of programmable hardware. For data center contexts, we also discuss application-aware monitoring and response as a potential mitigation.}

\section{Introduction}

The critical role of networks in our computing ecosystem has made them a high-value target for malicious actors, with high-profile attacks reported with alarming frequency and increasing severity. While the most common attacks target end systems or services like DNS, increasingly, networking devices themselves are coming under attack. The motivation is transparent: compromising network routers and switches provides visibility across large swathes of individual end devices. For many types of attackers, being able to compromise network devices is thus a higher priority~\cite{nsa-genie}.

While such attacks on network devices are not new, they have typically focused on either espionage or disruption. We explore a new type of threat
on the horizon, posed by an attacker who has obtained control over programmable switches in a target network: long-term deterioration in the network's service that is hard to diagnose. While it should be obvious that an attacker controlling powerful network devices can easily degrade performance to the point of denial-of-service, what distinguishes our work from well-known past attacks is the ability to operate while preventing diagnosis. Preventing or encumbering diagnosis is key to the attack's longevity; if the network operator could easily identify compromised devices, they would be removed or patched. We shall refer to such attacks, conducted with the objective of maximizing long-term, clearly visible damage while preventing diagnosis, as \attackName attacks.

A successful \attackName attack would result in long-term poor performance, and necessitate time-consuming manual intervention and analysis to eventually fix. Such attacks could be lucrative as ``ransomware'' -- attacking a popular public or private data center network, or ISP or enterprise network, in this way could cost the operator a substantial fraction of their revenue and hurt their reputation if not fixed, thus giving the attacker leverage. They could be broadly attractive to actors who today deploy denial-of-service attacks for ``hacktivism'', nuisance value, vendetta, etc.

Such attacks are non-trivial, even if the attacker controls powerful devices within a network. Network operators can deploy a variety of monitoring systems that aid in identifying anomalous network behavior and pinpointing faulty devices. 
Recent research, particularly in the data center context, has developed increasingly sophisticated methods to detect both total and partial failures, \eg Microsoft's 007~\cite{007}, Everflow~\cite{everflow}, NetBouncer~\cite{netbouncer}, and Pingmesh~\cite{pingmesh}, and a passive realtime monitoring system (henceforth, referred to as FB-mon) tested in a Facebook data center~\cite{passive-realtime}. These systems are a natural adversary for an \attackName attacker. 

We develop several \attackName attacks and evaluate their effectiveness in the face of a variety of monitoring and diagnosis approaches. We find that networked applications can be substantially degraded not only in most settings today, where monitoring systems are largely primitive, but also in the presence of industry-leading systems like 007, Everflow, and FB-mon. The success of such attacks stems from two observations. First, selectively targeting certain packets can cause disproportionate damage, while being invisible to most monitoring systems. At the transport layer, SYN packets and retransmissions, and at the application layer, messages part of a large ``coflow''~\cite{coflow}, are examples of such key packets. Second, programmable switches and routers, which are on the horizon, are highly effective for such targeting -- at switch fabric capacities of tens of Tbps, they can examine every single packet, and carefully select traffic to tamper with. Such hardware also enables limited \textit{stateful} processing, allowing complex operations across packet streams to increase the potency of \attackName attacks. This is a key point -- traditional switches, requiring switch CPU invocation for any sophisticated processing, can only manipulate three orders of magnitude less traffic, and lack the ability to carefully select this small sample of traffic.

Our work is based on the premise that network devices can be compromised, even inside environments secured through standard practices. As we discuss in some detail (\S\ref{sec:whyworry}), this founding premise is reasonable. Thus, our focus is not on exposing new vulnerabilities in network switches and routers, but on examining the potential impact of an attacker who has already gained control of such devices. We make the following contributions:
\begin{itemize}[leftmargin=5mm,itemsep=-1pt]
    \item We frame the notion of \attackName attacks, exploiting selective targeting of high-value packets using programmable switches, while preventing diagnosis. To our knowledge, this is the first consideration of such attacks.
    
    \item We show that transport-layer \attackName attacks can deteriorate performance by orders of magnitude for the targeted flows -- tens of seconds in the Internet context, and tens of milliseconds in data centers -- while not being amenable to diagnosis with today's monitoring systems. The use of programmable hardware vastly increases the potency of known and obvious attacks, while also enabling new ones.
    
    \item Similarly, attacks using application structure, such as that of large ``coflows'', can massively degrade application performance, causing, in some cases, $5\%$ of customer-facing requests to partially fail.
    
    \item We show that \attackName attacks can be implemented in P4, by storing a small amount of necessary state for target flows, and simple packet matching.
    
    \item Lastly, we formulate concrete recommendations for limiting the effectiveness of \attackName attacks.
    
\end{itemize}

\section{The risk of network device compromise}
\label{sec:whyworry}

Network devices, with their visibility across large numbers of end systems, are attractive targets for malicious actors. Several high-profile attacks on network infrastructure have come to light in recent times, using diverse attack vectors:

\begin{itemize}[leftmargin=5mm,itemsep=-1pt]
    \item More than $200$,$000$ Cisco switches deployed worldwide at ISPs, enterprises, and data centers were hacked in 2018 exploiting their management software~\cite{cisco200khacked}.
    \item A software vulnerability allowed the compromise of thousands of MikroTik routers~\cite{mikrotik-hacked} in 2018; such routers are often used in ISPs and enterprises~\cite{mikrotik-used}.
    \item US and UK governments issued a joint alert in 2018 warning of Russian hackers targeting ``routers, switches, firewalls and network intrusion detection systems at government and businesses''~\cite{russia-hacking}.
    \item In 2014, it was found that network equipment headed for deployment in many organizations was intercepted and modified with backdoors~\cite{nsa-device-intercept}.
    \item As part of a campaign to siphon data from an unnamed Central Asian government's data centers, it is suspected that an ISP's router was compromised~\cite{luckymouse} (March 2018).
    \item In 2013, the US intelligence agencies' ``Genie'' program came to light, that explicitly prioritized compromising network switches and routers~\cite{nsa-genie}.
\end{itemize}

\noindent Thus, for the foreseeable future, network devices are likely to remain a high-value target for actors ranging from state actors~\cite{russia-hacking, nsa-genie} to those focused on cryptomining~\cite{mikrotik-used}. 

Vulnerabilities in such devices are reportedly frequently; see for example, Cisco's weekly advisories~\cite{cisco-advisories}, which have disclosed several critical bugs, including in their data center network management software~\cite{cisco-advisory-CVE-2017-6639}, and in software in a popular line of switches~\cite{cisco-advisory-CVE-2016-1329}. Whitebox devices running Linux variants are also easy targets~\cite{defcon-whitebox}. Given that bugs have been found even in the simple tutorials for programmable switches~\cite{vera}, there is no reason to believe that such devices will be harder to compromise. 

In summary, while perimeter defense and hardening devices are useful and necessary measures, operators should also examine the question of ``What if an attacker does gain control over devices in my network?'' It is clear that industry leaders are actively considering this possibility even in facilities with best-in-class security:

\vspace{5pt}
\leftskip=0.3cm\rightskip=0.3cm\noindent\emph{``... it's not that a lot of people do encryption in the data center yet, but that's changing. We're doing that, among other reasons, because \textbf{if a malicious actor can break into one switch}, they can snoop on a lot of traffic ...''} 

\vspace{4pt}
\hspace*{\fill}--- Jitu Padhye, Microsoft~\cite{hotnets-dialogue-malicious}
\vspace{8pt}

\leftskip=0pt\rightskip=0pt

\noindent We examine the same possibility, of an attacker gaining control of network devices, in a different context. Instead of espionage, we focus on attacks where the motive is to degrade performance without allowing easy diagnosis, which would allow the operator to institute countermeasures that potentially neutralize the attack. Note that this is easier for espionage, than for \attackName's goal of actively degrading network performance such that it's impact is visible and obvious at the application layer.

\section{Attacker model}

\paragraphbb{Goals:} The \attackName attacker wants to cause large and obvious degradation in application performance, while making it difficult for the operator to localize compromised devices and perform diagnosis. More precisely, they want to increase flow completion time, while evading detection by state-of-the-art network monitoring systems.

\paragraphbb{Operating environments:} The attacker may sit within an ISP, or within a data center or enterprise network. In the ISP context, with data traveling across multiple autonomous systems, the goal is to avoid even localization to the ISP network in which the attacker's compromised devices reside. In the context of a data center or enterprise network, the goal is to leave open the possibility of external causes (\eg application layer problems, or problems rooted in external networks), or if the operator suspects the network, to slow or entirely prevent the pinpointing of compromised devices. The attacker's natural adversaries in both environments are network monitoring and diagnosis systems (\S\ref{sec:monSys}). In the data center setting, several sophisticated systems have been deployed or proposed, exploiting operator control over end hosts to achieve fine-grained monitoring. In the ISP context, in absence of such control, only more rudimentary methods are plausible.

\paragraphbb{Knowledge and capabilities:} We assume that the attacker:

\begin{enumerate}[topsep=0pt]
\setlength\itemsep{-3pt}
    \item has compromised some device(s) in a target network;
    \item these devices have programmable data planes (P4-like);
    \item knows the deployed monitoring system; and
    \item understands common transport, application primitives.
\end{enumerate}

Assumptions 1 and 2 are justified by attackers often succeeding in compromising network devices (\S\ref{sec:whyworry}), and the interest in deploying P4-like hardware. Assumption 3 is rooted in commonly deployed diagnosis systems being primitive, or/and publicly disclosed, together with the ability of the attacker to infer details of the monitoring system from observing the logging / sampling at their compromised devices. Assumption 4 reflects the ubiquity of TCP and common application patterns like ``coflows''~\cite{coflow}, which are sets of flows that accomplish an application-level objective.

\paragraphbb{Why does switch programmability matter?} Many of the attacks we shall discuss are either well-studied, or will be unsurprising for readers familiar with TCP semantics. The key point however, is that programmable hardware makes these attacks \emph{much} more potent: while a traditional switch CPU can only manipulate (very optimistically) tens of Gbps of traffic, a P4-like programmable switch, using its ASICs, can do so for the entire switch backplane, \eg $12.8$~Tbps today, \ie three orders of magnitude more traffic. The following operations can be performed at line-rate:

\begin{enumerate}[topsep=0pt]
\setlength\itemsep{-3pt}
    \item Reading \textit{all} packet headers, including matching against filters to select traffic for deeper processing.
    \item Stateful processing, whereby internal memory registers / counters can track flow state for a subset of traffic.
    \item Rewriting packet fields for a chosen subset of traffic, and generating packets by cloning benign ones, resubmitting them to the ingress pipeline, and rewriting them.
\end{enumerate}

\noindent We emphasize that while tampering with only a small amount of traffic is enough to seriously degrade performance, traditional switches lack the capability to carefully \textit{select} this small amount of traffic by examining \textit{all} traffic. Further, while coarse-grained manipulations are possible, \eg randomly sampling and dropping some traffic, these are either ineffective (at low sampling rate), or easy to detect.

\section{Monitoring systems as attacker adversaries}
\label{sec:monSys}

The \attackName attacker wants not only to maximize performance degradation, but also to evade detection. 
There is a large body of work on detecting faulty devices and anomalous network behavior, ranging from the simplest sampling and probing techniques to more recent work focused on detecting ``gray'' failures, where devices remain operational, but at degraded functionality~\cite{gray-achilles-heel, passive-realtime}. Detecting even benign gray failures is tricky because signals of such failure, such as packet loss and reduced bandwidth, are ambiguous, and can also result from normal congestion. Detection thus often rests on statistically separating normal and abnormal behavior by processing large amounts of network traces. We briefly describe several such systems, which will later serve as the benchmark for the undetectability of attacks. The more capable systems obviously are more limiting for the attacker.

\begin{table*}
\begin{threeparttable}
\begin{adjustbox}{max width=\linewidth}
\renewcommand*{\arraystretch}{0.5}
\begin{tabular}{c c c c c c c c c} 
& \multicolumn{7}{c}{\textbf{Monitoring Systems}} \\ 
\cmidrule(l){2-8} 
\textbf{Attacks} & sFlow & NetFlow & 007 & NetBouncer & Pingmesh & FB-Mon & Everflow & \textbf{Damage}  \\
\midrule \\
SYN Drop & \cmark  & \cmark & \cmark & \cmark & \cmark & \cmark \tnote{a,e} & \xmark & 64x Conn. Est. Time \tnote{c}\\ \\
Same Sequence Drop & \cmark & \cmark & \cmark  \tnote{d} & \cmark & \cmark   & \cmark  \tnote{a,e} & \cmark \tnote{b} & 50x FCT \tnote{c}\\ \\
SYN Flood & \cmark & \cmark & \cmark & \cmark & \cmark & \cmark \tnote{a,e} & \xmark & DoS\\ \\
RST Tinker & \cmark & \cmark & \cmark & \cmark & \cmark & \cmark \tnote{a,e} & \xmark & Disconnect\\ \\
ACK and Drop & \cmark & \cmark & \cmark \tnote{d} & \cmark & \cmark  & \cmark \tnote{a,e} & \cmark \tnote{b} & Disconnect \\ \\
ECN Tinker & \cmark & \cmark & \cmark & \cmark & \cmark & \cmark \tnote{a,e} & \cmark \tnote{b} & 6x FCT\\ \\
CWND Tinker & \cmark & \cmark & \cmark & \cmark & \cmark & \cmark \tnote{a,e} & \cmark \tnote{b} & 245x FCT\\ \\
Coordinated Incast & \cmark & \cmark & \cmark \tnote{a} & \cmark \tnote{a} & \cmark & \cmark \tnote{a} & \cmark \tnote{a} & 1.5x FCT\\ \\
\bottomrule 
\end{tabular}
\end{adjustbox}
\begin{tablenotes}
 \footnotesize
     \item[a] Using misdirection such that victim devices appear to be the faulty / compromised ones.
     \item[b] Everflow would require mirroring an infeasibly large amount of traffic to detect this attack.
     \item[c] Flows can be forced to time out as well.
     \item[d] Core switch can hurt all flows; ToR can only hurt upward inter-pod flows; agg. switch can hurt all intra-pod flows, and upward inter-pod flows.
     \item[e] Core switch can hurt all flows; all switches can hurt downward flows; agg. switch can hurt upward flows as well unless Core-IDs are later updated.
\end{tablenotes}
\caption{\attackName attacks are effective even with sophisticated monitoring systems in place.
}
\label{tab:monitoring}
\end{threeparttable}%
\end{table*}

\paragraphbb{Sampling and coarse statistics:} Widely deployed monitoring systems are largely primitive. For instance, ISPs commonly depend on sampling packets, typically at the rate of one in thousands~\cite{sflowsamplingrates}, using tools like \textit{NetFlow}~\cite{netflow} and \textit{sFlow}~\cite{sflow}. Such solutions under-represent small flows, and can't track packets through the network, making even simple metrics like drop rate hard to infer. These are easy targets for \attackName attackers. It is likely that similar systems are in use at enterprises and data centers, except at the industry leaders like Google, Microsoft, Facebook, etc.

\paragraphbb{Targeted mirroring:} One can use simple packet matching rules to mirror packets, \eg TCP control packets, to a central controller using \textit{Everflow}~\cite{everflow}. With suitable mirroring rules, this can impair several easy attacks that target packets like SYNs, as discrepancies in mirrored packets across switches on a path will be obvious.

\paragraphbb{Active probing:} A data center operator can leverage control over endhosts to probe end-end paths. For instance, \textit{Pingmesh}~\cite{pingmesh} generates TCP-ping probes between endhosts using different port numbers to explore different path, and can identify paths with poor latency or excessive loss of probes. \textit{007}~\cite{007}, instead of using active probing continually, initiates traceroute probes when it detects retransmissions at endhosts, and uses a tomography technique across these probes to identify faulty links. Another similar approach is \textit{NetBouncer}~\cite{netbouncer}, which uses IP-in-IP tunneling to send probes to core switches that sit at the top of a Clos network hierarchy, ensuring a unique path is tested with each probe. Paths where probes are lost are identified as congested.

\paragraphbb{TCP statistics on live traffic:}
Instead of generating probe traffic, an operator can also use TCP statistics across real production traffic to detect problems. \textit{FB-mon}~\cite{passive-realtime} uses this approach, drawing on the assumption that the distribution of characteristics like latency and retransmissions across different paths should be similar in a highly symmetrical data center network. Thus paths that deviate substantially from these transport-level outcomes for live traffic can be flagged. FB-mon relies on switches to identify the paths packets take.

While there are many other research proposals promising per-packet or per-flow visibility~\cite{dapper, flowradar, turboflow}, these need similar mechanisms as the above broad taxonomy of systems to distinguish anomalous behavior from normal congestion.
Researchers have also explicitly focused on detecting malicious devices~\cite{mizrak-malicious-routers, mizrak-packet-loss}, by modeling the expected input-output behavior and queueing in the devices, and verifying that traffic observations fit these models. It is unclear whether such details can be captured for large modern devices, with high port counts, line rates, and shared-buffers. 

Note also that indeed, programmable hardware will enable superior monitoring systems, but this does not necessarily ameliorate \attackName attacks. The potency of such attacks stems from the vastly different burdens of the monitoring systems and the attackers: the attacker must only do stateful processing over a small fraction of traffic, while a monitoring system must cover virtually all traffic to ensure some fraction of it is not tampered with.

\section{Transport and network layer attacks}
\label{sec:transport-layer-network-layer-attacks}

\begin{figure}
\centering
\includegraphics[width=\columnwidth]{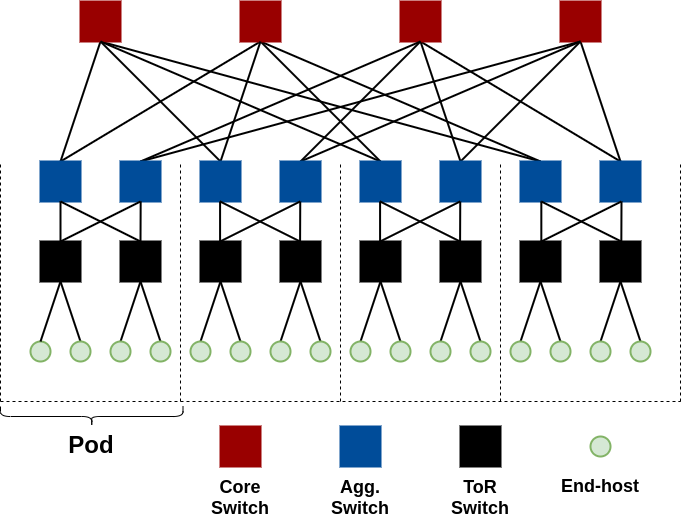}
\caption{A small, k-4 (\ie each switch has $4$ ports) fat-tree network used as a reference. This type of topologies are common in deployed data centers~\cite{jupiter-rising}. The top layer of switches are referred to as ``core'', middle as ``aggregation'', and the lowest layer switches that are directly attached to servers, are top-of-rack (ToR) switches.}
\label{fig:k4fattree_benchmark_diagram}
\end{figure}

We discuss several transport and network layer \attackName attacks, and evaluate their effectiveness in the presence of a range of monitoring systems (\S\ref{sec:monSys}). Since the more sophisticated monitoring systems of the data center setting are more challenging, we present results in the context of a standard fat-tree data center network~\cite{fattree}, illustrated in Fig.~\ref{fig:k4fattree_benchmark_diagram}. We evaluate two types of attacks in different emulation environments: (a) single-switch attacks, using the behavioral model for P4 switches (P4-BMV2) integrated into Mininet~\cite{bmv2}; and (b) coordinated attacks from multiple compromised switches, using ns-3~\cite{ns3} with ECMP support~\cite{kheirkhah2016mmptcp}. Table~\ref{tab:monitoring} summarizes our results.

\subsection{SYN drop} 
\begin{figure}
     \centering
      \includegraphics[width=0.7\columnwidth]{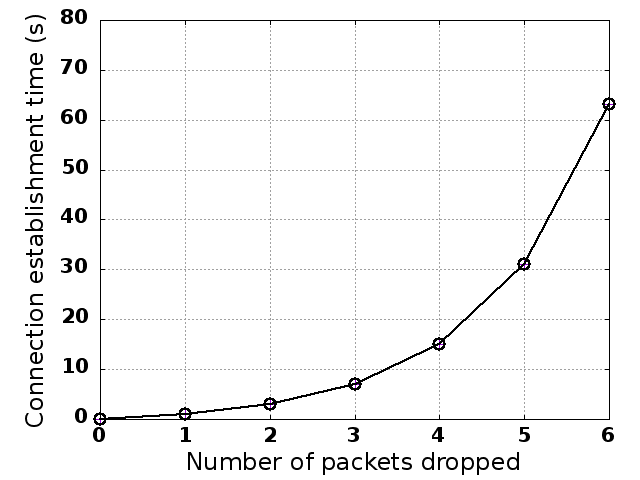}
        
        \caption{The impact of repeatedly dropping a SYN. These experiments are implemented in Mininet/P4-BMV2, and use a $5$~Mbps, $2$~ms RTT connection, with a $200$~ms min. retransmission timeout (Linux default). With lower timeout configurations in data centers, the impact will look similar, but with lower absolute degradation.} 
        \label{fig:syn_drop_general}
\end{figure}

The attacker repeatedly drops the SYN packets of a target flow; after each drop, the client incurs exponential backoff before retransmission. Connection establishment time can thus be increased exponentially until the network stack's configured number of retries. Fig.~\ref{fig:syn_drop_general} shows this experimentally using our implementation in P4-BMV2.

This obvious attack illustrates the utility of programmable hardware: the compromised switch can store SYN packet headers for targeted flows in a bloom filter, match all future traffic against this filter, and drop SYNs for target flows. This type of SYN matching at line-rate has already been successfully demonstrated on P4 hardware for other applications like stateful load balancing~\cite{silkroad}, but we nevertheless successfully implemented this attack in P4-BMV2. Note that even if a traditional switch could identify SYNs, repeated targeting of the same SYNs requires stateful per-packet processing beyond its capabilities. Dropping all or a large fraction of SYNs without careful targeting would be easy to detect with even naive sFlow-based monitoring.

The fraction of packets such an attack must target to severely impair performance is extremely small: a large fraction of SYNs typically comprise only a small fraction of traffic. Thus, naive sampling-based monitoring with sFlow/NetFlow is ineffective. Systems that use active probes to look for congestion (NetBouncer, Pingmesh) are similarly ineffective, especially if the attacker understands the monitoring system and does not impair the probes. 

\begin{figure}
\centering
\includegraphics[width=\columnwidth]{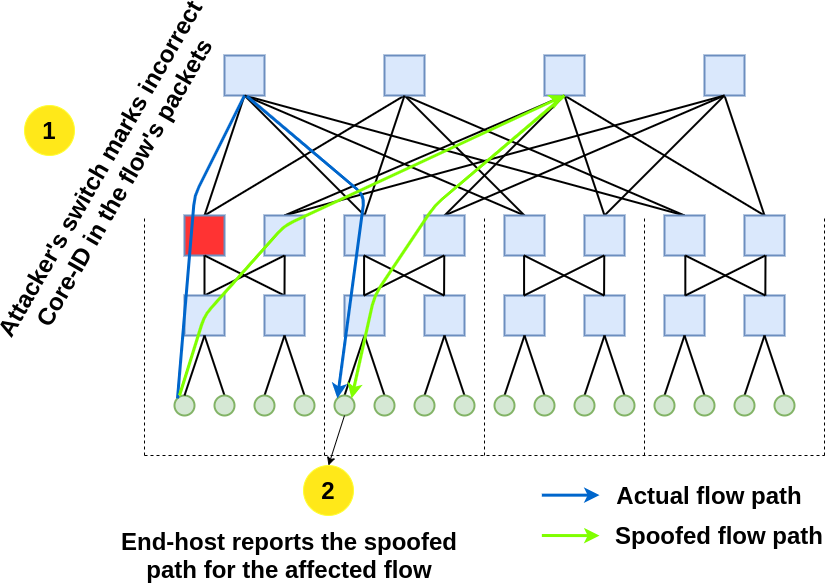}
\caption{To misdirect FB-mon, the attacker's compromised switch (marked in red) can mark incorrect core-IDs, thus, raising a false alarm at some other switch. If the core switch is responsible for core-ID marking, an aggregation switch may only hurt downward flows by rewriting their core-IDs.}
\label{fig:k4fattree_singleattacker_fbmon}
\end{figure}
\begin{figure}
\centering
\includegraphics[width=\columnwidth]{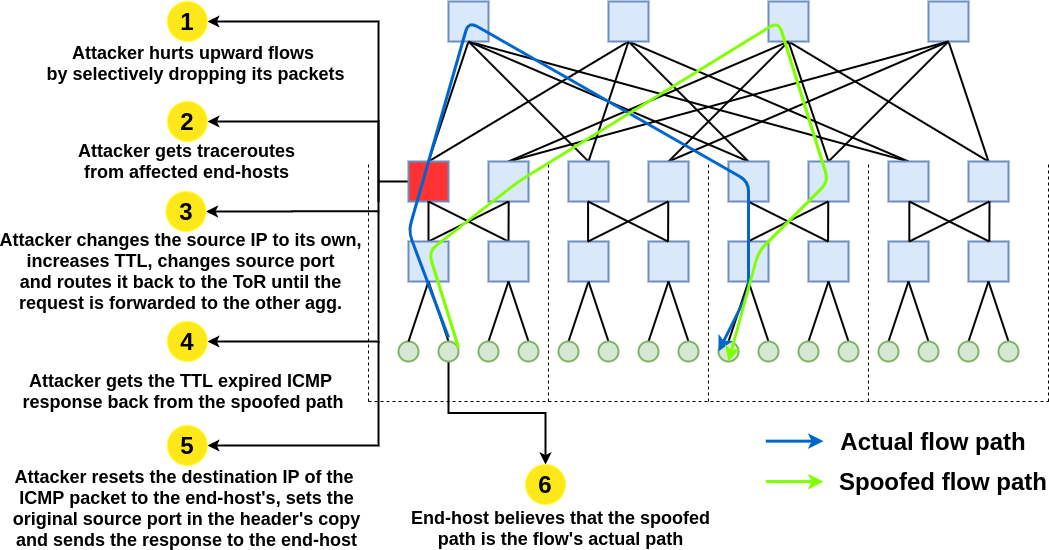}
\caption{To misdirect 007, a malicious switch (marked in red) discovers a fake traceroute path for the end-host. This can be accomplished even when the attacker does not have any knowledge about IPs assigned to other switches.}
\label{fig:k4fattree_singleattacker_007}
\end{figure}

FB-mon will indeed raise alerts about poor observed performance from TCP statistics in live traffic. However, the attacker can, in many cases, avoid detection by casting blame on other switches. FB-mon relies on on-path switches to identify paths by marking which core switch (``core-ID'') was used. Compromised core switches and aggregation switches can simply mark the wrong core-IDs in packets, thus leading FB-mon to flag the wrong paths / devices as faulty. This is illustrated in Fig.~\ref{fig:k4fattree_singleattacker_fbmon}.

007 sends traceroutes on seeing losses and could potentially result in a large number of traceroutes passing through the compromised switch, thus flagging it. However, detection can be avoided by tampering with the ICMP traceroute probes. Traceroutes are easy to detect because of their TTL fields. The compromised switch can generate ICMP responses with IP addresses of other switches on other paths. This can be done such that the end-host receives a series of traceroute responses corresponding to other paths in the network. If IP addressing for other switches is unknown, this requires more effort, but is nevertheless possible: the attacker can change the source address field in the ICMP requests to its own and get responses from other switches, then forward them to the traceroute source with appropriate TTLs. The full workings of this process (without the attacker knowing datacenter-wide IP addressing) are illustrated step-by-step in Fig.~\ref{fig:k4fattree_singleattacker_007}.

The only system that fully prevents this attack is Everflow, which mirrors all SYN packets, and would quickly flag discrepancies in mirrored SYNs across the compromised device(s) and their immediately neighboring devices.

\subsection{Same sequence drop} 
\begin{figure}
     \centering
      \includegraphics[width=\columnwidth]{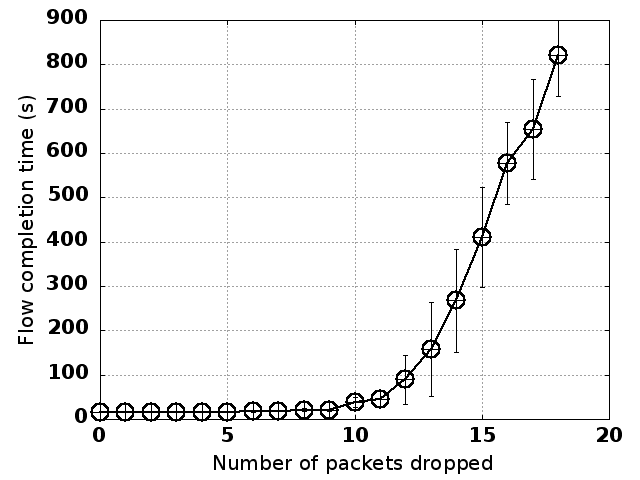}
        \caption{The impact of repeatedly dropping the same sequence number packet. The setup is the same as for the SYN-drop experiments. The variation for sequence drop stems from varying fast retransmit behavior.}
        \label{fig:sameseqdrop_sub}
\end{figure}

This is a similar attack to the above, except, instead of targeting SYNs, the compromised switch repeatedly drops packets with the same sequence number, thus preventing a TCP flow from making progress. The attack mechanics, impact, and response of monitoring systems is broadly similar to the SYN drop attack. Fig.~\ref{fig:sameseqdrop_sub} shows the results from P4-BMV2.

Prior work~\cite{lowrateshrew} explored a similar attack using malicious end-hosts that target a victim sender by pushing coordinated traffic bursts. The bursts try to synchronize in a heuristic manner to arrive at the same time as the victim's retransmissions. An attacker's control over on-path programmable devices makes this imprecise approach unnecessary. 

\subsection{SYN flood} 

This standard denial-of-service approach can be made much more effective with compromised programmable switches. The attacker generates a stream of SYNs with different source IP addresses targeting a victim end-host. The victim can ameliorate this traditional attack using SYN cookies, whereby it responds with a SYN-ACK, but does not allocate state for the connection until receiving the cookie back~\cite{rfc4987synflood}. However, a P4-switch that is appropriately located to receive the SYN-ACKs (\eg a ToR switch to which the victim server is attached) can easily generate packets responding to the cookie: this requires incrementing the sequence number, which is trivial at line-rate for P4 hardware.

A stream of SYNs used in this way, embedded in a large amount of traffic, is unlikely to be detected by random sampling. It is also unlikely to cause congestion in the network, rather only exhausting the server's SYN queue to block new connections. Thus, congestion-detecting systems like 007, NetBouncer, and Pingmesh are ineffective. FB-mon will observe connections being denied by the victim server, but as these incoming connections traverse the entire network, it can be tricked into flagging the wrong paths, as noted above.

Everflow, by mirroring all SYNs, will see that neighboring benign switches do not have the same SYN stream, and thus flag the compromised switch.

\subsection{RST tinker} 

A P4 switch can easily mark the reset (RST) bit of any \textit{one} packet in a flow, causing disconnection. Monitoring systems looking for loss or congestion are ineffective against this attack. 007 does not get an opportunity to launch traceroutes, as the flow terminates abruptly instead of continuing despite losses. FB-mon will observe many failed connections, but can be misdirected to blame other paths / devices. Everflow, by mirroring control packets, including RSTs, will catch this attack as well.

\subsection{ACK and drop} 

A similar effect can be achieved by the P4 switch dropping a single arbitrary packet, but generating and sending an ACK for it to the sender. This causes the sender to slide their sending window forward, while the receiver awaits the dropped segment. Experimentally, we observed that unlike a RST's quick termination, the connection continues with zero throughput for nearly $1000$~seconds. Pingmesh, NetBouncer, sFlow, Netflow are ineffective here; 007 and FB-mon will observe packet drops and poor TCP performance respectively, but can both be misdirected as discussed earlier. Even Everflow will only catch this attack if all retransmissions are being mirrored, or with manual intervention to set debug bits for packets in victim connections. Even such monitoring can be thwarted by the attacker by ensuring the debug packets are not tampered with.

\begin{figure}
     \centering
         \includegraphics[width=\columnwidth]{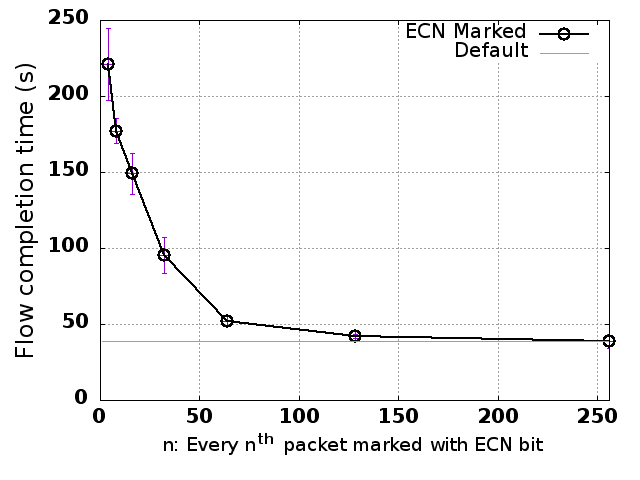}
        \caption{Marking even a small fraction of packets with ECN bits causes a large deterioration in FCT.}
         \label{fig:ecn_sub}
\end{figure}

\begin{figure}
     \centering
         \includegraphics[width=\columnwidth]{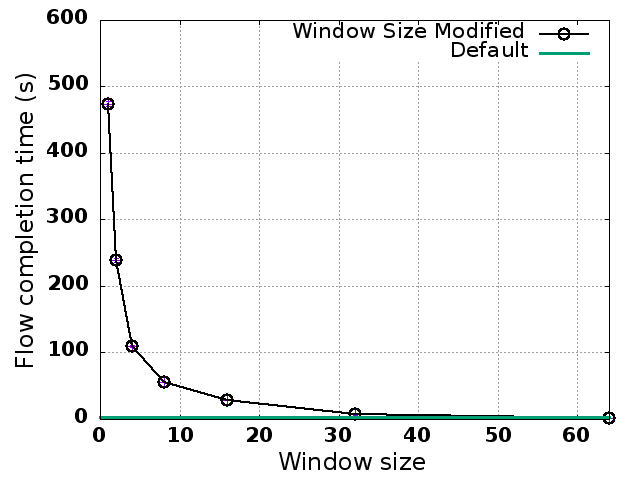}
         \caption{Rewriting the window size to smaller values impairs performance.}
         \label{fig:cwnd_long_sub}
\end{figure}

\subsection{ECN tinker} 

Explicit Congestion Notification bits in packets (echoed in ACKs) are used by network switches to signal high queue occupancy and congestion to senders. Senders who receive ECN-marked packets cut down their sending rate in response. Thus, a P4 switch can mark ECN bits for packets in a flow to deteriorate its performance even in the absence of congestion. The degree of damage done can be controlled by the attacker by deciding how many packets to mark. Fig.~\ref{fig:ecn_sub} shows, using P4-BMV2 experiments, the deterioration in flow completion time (FCT) with different degrees of ECN markings. Marking only $3\%$ of segments results in an FCT increase of more than $2\times$.

Low-rate marking for a limited set of target flows will be unlikely to reveal significantly different rates of marking in coarse packet samples with sFlow/Netflow. This also does not cause congestion for other flows, and will not be noticed by 007, NetBouncer, and Pingmesh. FB-mon will observe the poor TCP performance for paths traversing compromised switches, but can be misdirected to place blame on benign switches, as noted earlier. The analysis for Everflow is similar to the above ACK-and-drop attack.

\subsection{CWND tinker} 

A TCP sender's sending window size determines their throughput. It is adjusted in response to network congestion, and the receiver's advertised receive window. An on-path P4 switch modifying the receive window carried in ACKs thus forces the sender to reduce their rate. Fig.~\ref{fig:cwnd_long_sub} shows this attack with different arbitrary receive window sizes. A large deterioration in FCT can be caused, \eg $25\times$ worse with window size of $8$, and $245\times$ worse with window size of $1$. (The window scaling factor is $512$.) This attack's effectiveness under different monitoring systems is similar to that of ECN tinker above.

\subsection{Coordinated incast} 
\label{subsection:incast}
An attacker who has compromised multiple switches can create routing imbalance in the network. Typically, routing in data centers uses ECMP, which uses hashes of packet headers to spread flows across different equal-length paths. This is essentially an implementation of randomized load balancing at flow granularity. This attack hinges on deliberately skewing this load balancing.

\begin{figure}
\centering
\includegraphics[width=\columnwidth]{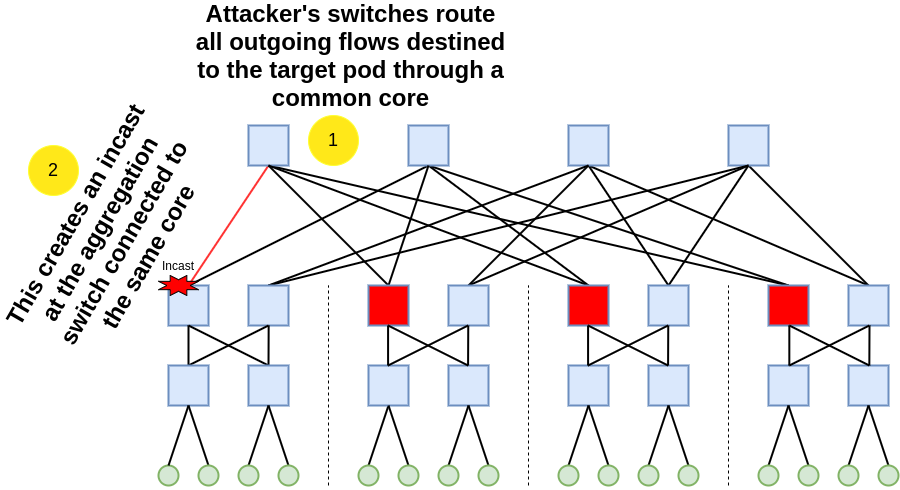}
\caption{A coordinated incast attack on a k-4 Fat-tree. Attackers (marked red / dark) forward all flows destined to the first pod through a common core switch which creates an incast-like situation at the down-link (marked red)}
\label{fig:k4fattree_incast_diagram}
\end{figure}

Multiple compromised switches choose a subset of traffic for which they pick the same route. An example of this is shown in Fig.~\ref{fig:k4fattree_incast_diagram}, where three compromised switches send all traffic destined to the left pod through the leftmost core switch, instead of balancing it across all cores. This congests the downlink(s) from that core switch to the pod. Malicious switches also do not need to continually act in this manner; they can route normally (using ECMP) most of the time, and still cause substantial damage by occasional coordinated imbalance.

To evaluate this attack's effectiveness, we use a packet-level simulation in ns-3, emulating a fat-tree with $45$ switches (larger than Fig.~\ref{fig:k4fattree_incast_diagram}). We use the a Web search workload from prior work~\cite{pfabric} to draw flows of different sizes, and vary their (Poisson) arrival rate. We assume $5$ switches to be compromised and causing imbalance towards one pod (similarly to Fig.~\ref{fig:k4fattree_incast_diagram}). We compare the FCTs at different loads for different degrees of traffic skew caused by the malicious switches.

\begin{figure}
\centering
\includegraphics[width=\columnwidth]{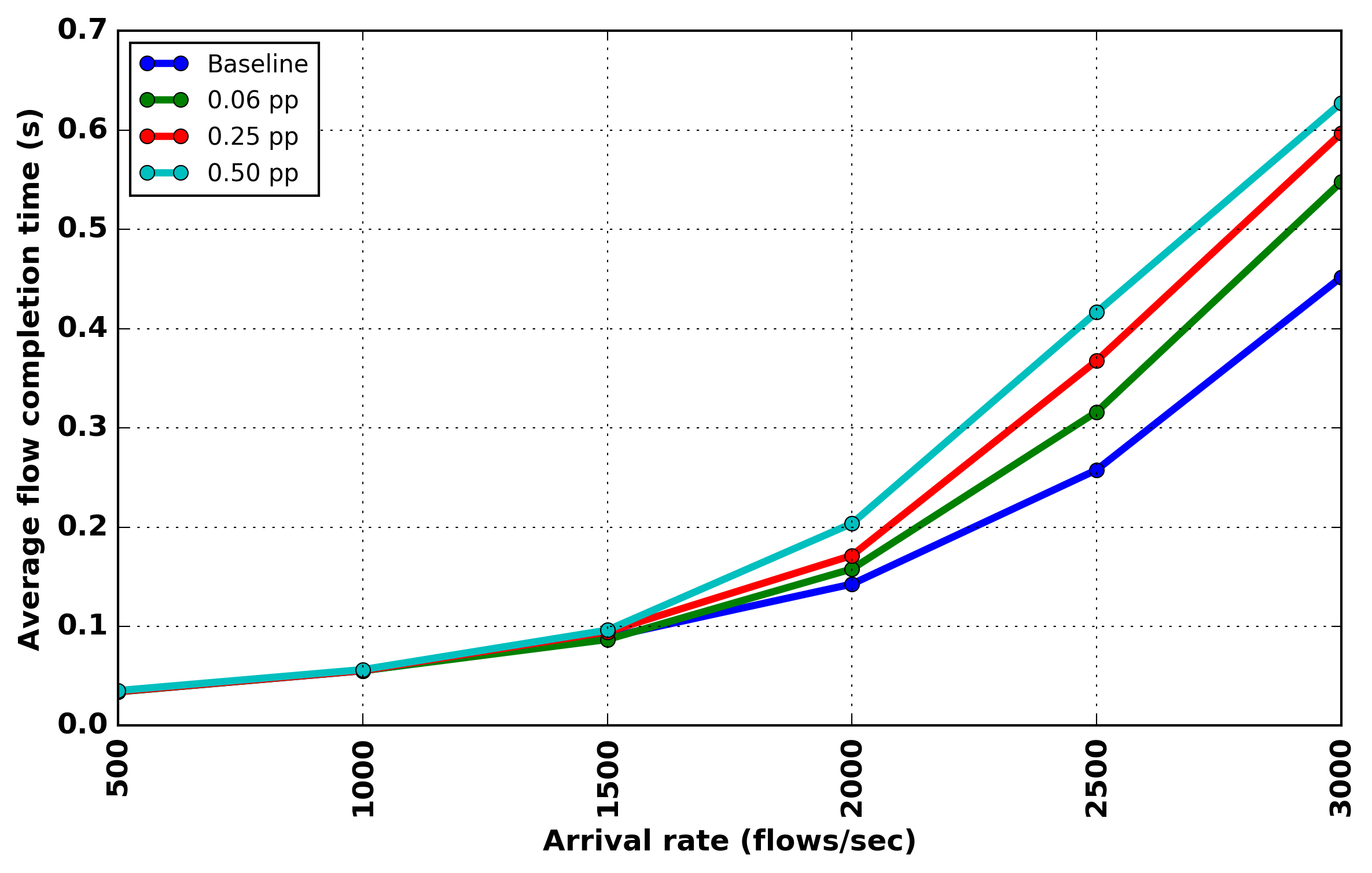}
\caption{Average flow completion time analysis for flows destined to the victim pod in a coordinated incast attack.}
\label{fig:k6fattree_incast}
\end{figure}

Fig.~\ref{fig:k6fattree_incast} shows the increase in FCT (for flows destined to the left pod) as compromised switches increase their skew for precisely the topology and compromised switch setting shown in Fig.~\ref{fig:k4fattree_incast_diagram}. A ``pulse proportion'', pp=$0.06$ implies normal operation $94\%$ of the time, with only $6\%$ time spent diverting traffic for the left pod to one core switch. At low load, the network still has enough capacity for FCT to not increase, but as load increases, poor load balancing hurts FCT. 

An interesting feature of this attack is that it creates congestion and performance deterioration far from the malicious switches. Monitoring systems can thus only identify the remote, affected location. The compromised switches can also use misdirection (like in some of the other attacks above) to further cloud diagnosis. 

\subsection{Summary and other attacks} 

Referring back to Table~\ref{tab:monitoring}, we see that we can deteriorate performance for targeted flows by several times, in the worst case leading to complete disconnection and denial of service. The demonstrated attacks require only simple, easy-with-P4, line-rate manipulation of a small fraction of packets for target flows; in many cases, tampering with only one packet in a target flow (\eg RST tinker, ACK and drop) suffices. Also note that the attacker can flexibly choose target traffic, deciding to deteriorate performance for specific destinations, or in a stochastic manner for some fraction of flows or destinations. The most commonly used monitoring approaches, based on random sampling, will not even raise alerts (let alone pinpoint malicious devices) unless the attacker targets such large fractions of traffic that aggregate metrics like total traffic volume or loss-rate change substantially. Industry-leading monitoring systems will, in some cases, raise alerts about anomalous behavior, but can be easily misdirected such that the operator assigns blame to benign devices, instead of being able to pinpoint malicious ones. Note that targeting a small but time-variant set of destinations over time can cause massive damage to a provider's reputation --- several different customers will observe severe impairments at different points in time, with no easy, available recourse.

We only discussed a sampling of transport and network layer attacks, but there are, of course, many other possibilities, \eg generating duplicate ACKs from which a sender infers loss and congestion, thus reducing its rate; corrupting the packet payload and updating the checksum to match it, such that transport silently delivers corrupt data to the application; bouncing packets between multiple compromised switches to increase their latency. We expect many more creative attacks to be put forth over time; our goal is only to illustrate that programmable hardware (a) makes simple, expected attacks more effective; and (b) enables entirely new types of attacks like the coordinated incast.

\paragraphbb{The Internet context} is much easier than data centers, allowing for even more successful \attackName attacks for several reasons: (a) not having end-host support (unlike data centers) limits the design of monitoring systems; (b) the operator does not have an end-to-end view of the network path, making it is easy to blame any problems on external networks; and (c) round-trip times across the network being longer increase the degree of damage from several attacks, \eg SYN or sequence number drops.
\section{Application-aware \attackName attacks}
\label{sec:application-aware-attacks}

While hampering specific target flows is interesting, the impact of such attacks can be amplified if the attacker also exploits knowledge of application structure. Even if sophisticated monitoring systems were able to catch any attacks that impact more than a small threshold fraction of traffic, using application-level knowledge to carefully select target traffic within this bound can still allow \attackName attacks to be potent.

Many applications in such environments have a large dependency set, and slowing communication between components can slow down the application. This means that network flows are often part of a multi-flow application-level task: a ``coflow''~\cite{coflow}. For instance, MapReduce jobs result in data shuffles whereby the map tasks communicate with the reduce tasks, and the job awaits the termination of the last reducer to finish or to start the next iteration. Thus, the progress of compute tasks and the freeing of their resources necessitates that the last communication flows finish quickly. Similarly, user-level responses in Web applications (\eg search results) often use data requested from hundreds~\cite{facebook-memcache} to thousands~\cite{thousandCoflows} of components in a ``partition aggregate'' manner. Given that user-experience in terms of response time is critical to the application providers' revenue and reputation, it is desirable that such coflows (and hence, the last flows within them) finish as quickly as possible.

\begin{figure}
\centering
\includegraphics[width=8cm]{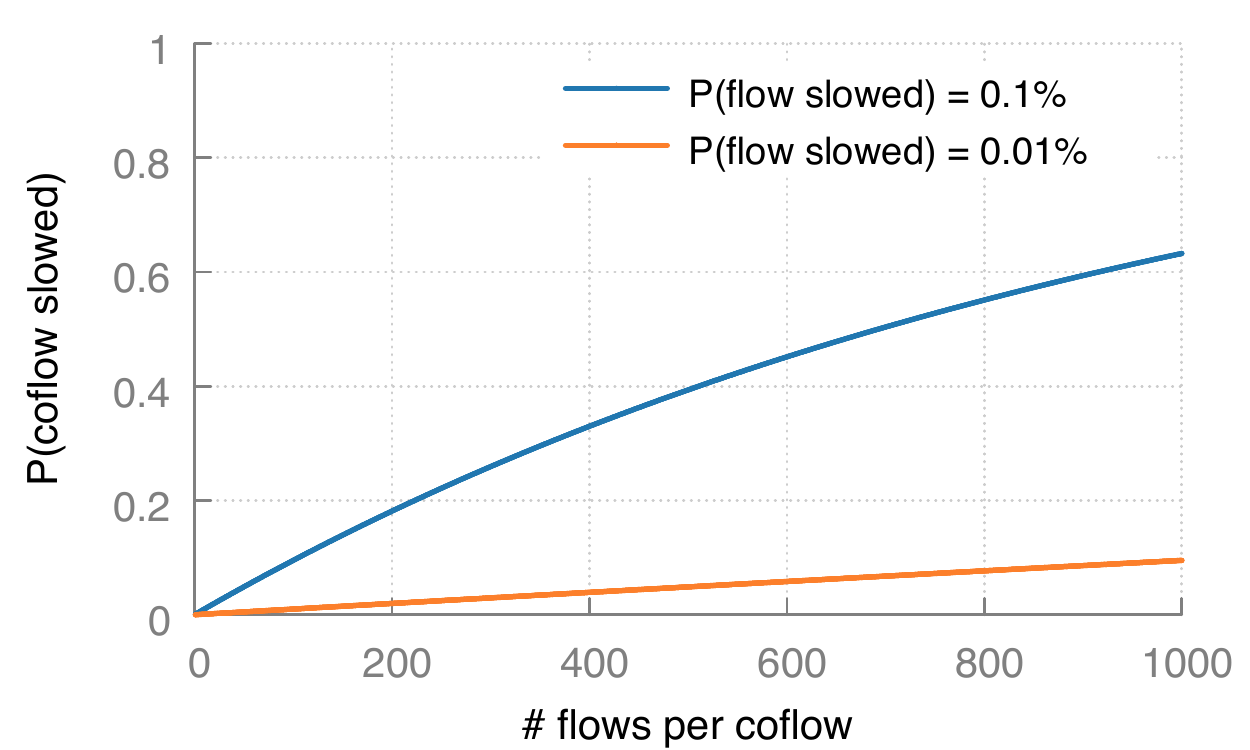}
\caption{In the face of dependencies, a minor probability of a single flow being slowed can yield a major probability of the spanning coflow being slowed.}
\label{fig:bin-coflow-slowdown}
\end{figure}

The completion of such coflows being dependent on the last few straggler flows opens up substantial opportunity for an attacker to degrade performance for a large number of coflows while targeting only a small set of flows. An illustration of this effect is shown in Fig.~\ref{fig:bin-coflow-slowdown}. Even if the attacker is constrained by advanced monitoring systems to only cause a slowdown for less than $0.1\%$ of flows, if each coflow comprises $1000$ flows that depend on the last finishing flow, fully $63\%$ of coflows experience slowdown\footnote{This effect is well-studied in prior work~\cite{tail-at-scale} in the context of failures; our contribution here is in exploiting this natural characteristic of such systems from an attacker's perspective.}. Thus, monitoring systems in networks under such attacks may find that only an insignificant number of flows are impacted, a situation largely indistinguishable from normal congestion. But at the application layer, the large impact would be clearly visible. This is likely to cause a wild goose chase into the myriad of \textit{other} possible causes such as application software bugs, misconfiguration, resource contention, or faulty hardware. The problem could be exacerbated if the network operator and the application owner are different parties, as is the case in popular cloud infrastructure providers with their customers.

How much the slowdown of constituent flows slows down coflows depends on application structure. For example, MapReduce has a hard barrier which causes the slowest flow to determine the coflow completion time. On the other hand, a Web page could potentially be partially returned and still be useful enough. As we shall see later, techniques for straggler mitigation, such as request ``hedging''~\cite{tail-at-scale}, can further soften this dependence, and weaken an attacker. 

To illustrate the capabilities of an attacker in such a setting, we simplify the context to a minimal example. Nevertheless, such application structures as modeled, are common in use today.

\paragraphbb{Application model.} We will assume we have $n$ uni-directional coflows going through a single switch, \eg a top-of-rack switch which hosts the server responsible for user-level jobs or queries. The failure of user-level requests ("coflows") is assumed to impact revenue linearly. (This is likely to be a conservative assumption.) Even failures of small fractions ($1$-$5\%$) of user queries is disastrous: imagine Amazon's customers experiencing a $1\%$ probability of experiencing a timeout (or missing thumbnails, etc.) on pages they want to load. 

Each coflow is modeled as a fan out of $m$ requests to different backend servers going over this switch; $m=500$ is close to the $521$ (average) reported by Facebook for building a popular user-facing page \cite{facebook-memcache}, but Web search backends can involve $m$ values in thousands~\cite{thousandCoflows}. For simplicity, each request is assumed to be a single-packet UDP flow. We vary the number $F=\{1, 2, 5, 10\}$ of the $m$ requests that have to fail for a coflow to fail.

\paragraphbb{Hedging.} With $500$ requests and the failure of only one being enough for the coflow to fail, \ie $m=500$ and $F=1$, even a random drop rate of merely $0.01\%$ will cause $4.9\%$ coflows to fail. This is already unacceptably high, so a request replication technique is commonly used for such applications, called ``hedging''. For each request sent out, we sent out an additional redundant request, so that the replication factor, $r=2$. Such hedging's use at Google, together with techniques that eliminate most of its overhead, has been detailed in prior work~\cite{tail-at-scale}. Hedging is extremely effective in addressing random drops, leading to a nearly $0\%$ coflow failure rate, because both the original query and its replica query must incur a random drop for the query (and the coflow) to fail.

\paragraphbb{Misclassification model.} To be effective, an attacker needs to be able to classify constituent flows into their coflows, and thus track hedged flows. A request packet is identified by the tuple $(id_{coflow}\in\{1, ..., n\}, id_{req}\in\{1, ..., m\})$. The classification ability of the attacker, who has obtained control of the switch, is modeled as a Bernoulli distribution with $p_{mc}$, the probability of \emph{misclassification}. Thus, the attacker can successfully map a request to its request and coflow identifiers with probability $1 - p_{mc}$. Otherwise, the desired observation of the real $(id_{coflow}, id_{req})$ is replaced by a uniform random pick.

\paragraphbb{Budget model.} For illustration, we use an overall packet drop rate budget $D=0.01\%$, which is lower than the detection threshold of state-of-the-art monitoring systems. The total drop budget across all $n$ coflows is then $B=n\cdot r\cdot m\cdot D$ packets. Without hedging, the budget is thus $0.05n$ and with hedging, $0.1n$. Per coflow, this translates (with $D=0.01\%$, $m=500$) to $0.05$ or $0.1$ packets allowed to be dropped on average. With perfect knowledge about which flows belong to which coflows and which requests are hedged versions of which others (\ie matching request identifiers), we could fail $5\%$ of the coflows (with or without hedging) by dropping only 1 in 10000 packets ($D=0.01\%$).

\begin{figure}
\centering
\includegraphics[width=8cm]{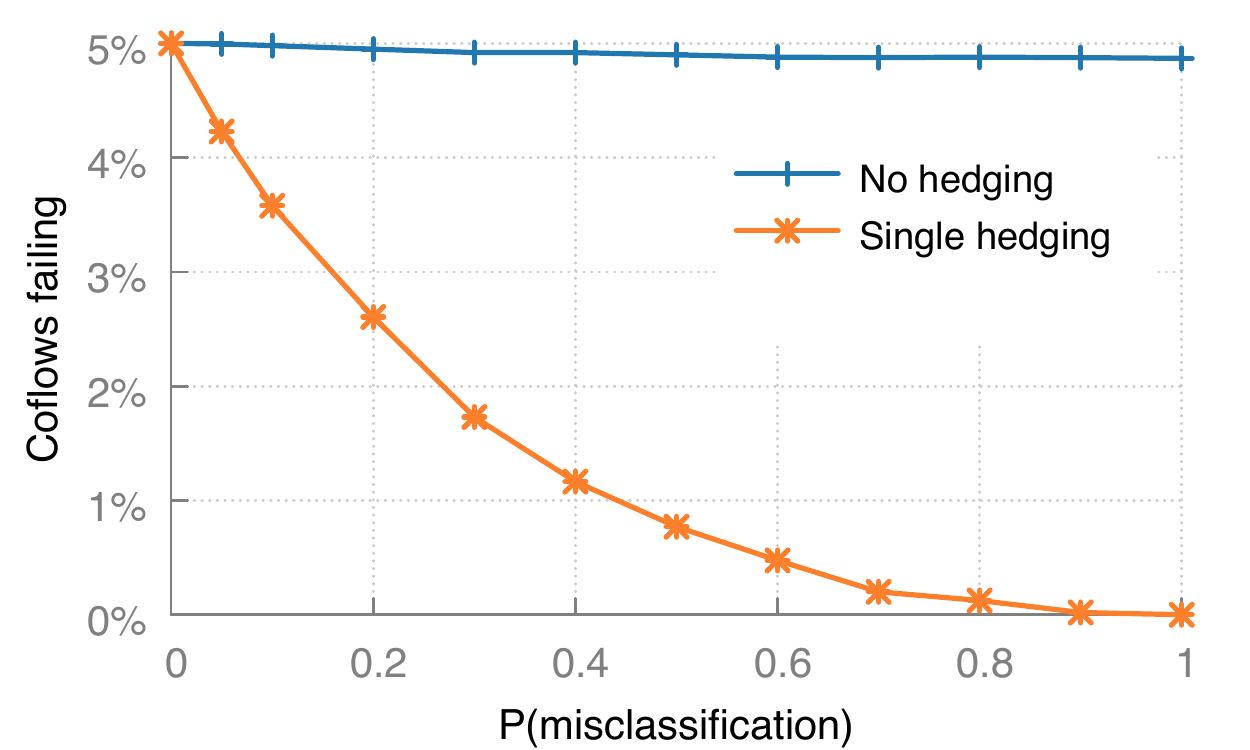}
\caption{Facing imperfect knowledge, hedging reduces the ability to fail coflows ($m=500$, $F=1$, $B=0.01\%$).}
\label{fig:hedging-requests}
\end{figure}

\paragraphbb{What is the effect of misclassification?} We vary the $p_{mc}\in[0, 1]$ with or without hedging, assuming $F=1$ request must fail to fail the entire coflow (shown in Fig. \ref{fig:hedging-requests}). Without hedging, even random loss will cause coflows to fail ($4.87\%$, $\sigma=0.0005$) at a nearly equal rate to perfect knowledge (5\%, $\sigma=0$). The small difference stems from multiple requests from the same flow being dropped on occasion, leading to no additional negative impact. With hedging, the reduced ability to classify requests causes significantly fewer coflows to fail. With completely random loss ($p_{mc}=1$), zero coflows failed in our tests, confirming analysis in previous work~\cite{tail-at-scale} showing the effectiveness of hedging against random loss. However, as the attacker becomes more capable, \ie $p_{mc}$ decreases, hedging becomes less effective; an attacker with $55\%$ classification accuracy ($p_{mc}=0.45$) can cause $1\%$ of coflows to fail.

\begin{figure}
\centering
\includegraphics[width=8cm]{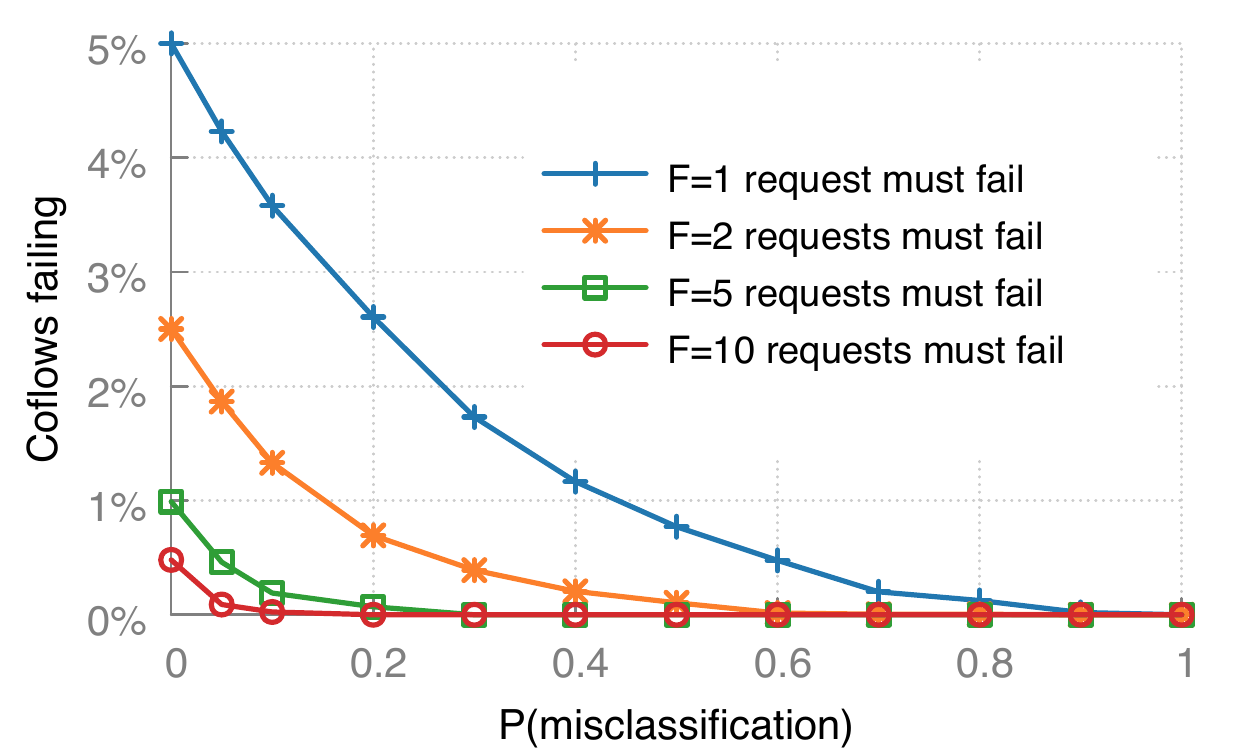}
\caption{The more requests must fail ($F$) for the coflow to fail, the less coflows fail ($m=500$, $B=0.01\%$; single hedging).}
\label{fig:request-for-coflow-fail}
\end{figure}

\paragraphbb{How vulnerable can the coflow be?} We vary the $p_{mc}\in[0, 1]$ with hedging, for a range of $F=\{1, 2, 5, 10\}$ requests needing to be failed to fail the entire coflow (shown in Fig. \ref{fig:request-for-coflow-fail}). The more requests that must fail to fail the coflow, the less coflows will fail. This is caused by the increased budget spent to fail the necessary amount of requests. By increasing $F$, the amplification effect of the attack is reduced. Because $F$ requests must fail, it is necessary for the attacker to drop at least $F$ hedged requests, which becomes exponentially more difficult as the probability to classify correctly decreases linearly. Typically it is not possible to make the coflow more or less vulnerable, as this is decided by the application's nature.

\begin{figure}
\centering
\includegraphics[width=8cm]{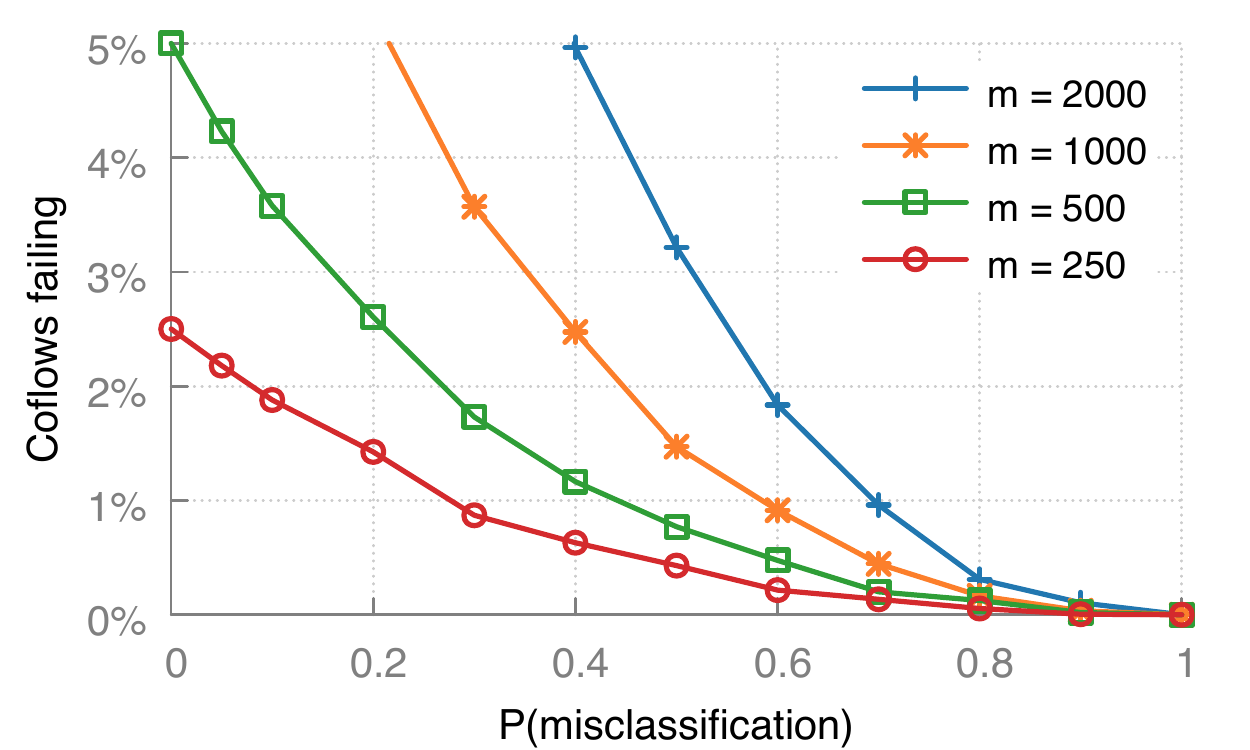}
\caption{The smaller the fan-out ($m$), the less coflows are able to be failed ($F=1$, $B=0.01\%$; single hedging).}
\label{fig:fan-out}
\end{figure}

\paragraphbb{What is the effect of fan-out?} We vary the $p_{mc}\in[0, 1]$ with hedging and $F=1$, for a range of $m\in\{250, 500, 1000, 2000\}$ requests are sent (shown in Fig. \ref{fig:fan-out}). With a linear decrease in fan-out, the available budget also decreases linearly. It is important to note that for $m=250$ in Fig. \ref{fig:fan-out} is always higher than $F=2$ in Fig. \ref{fig:request-for-coflow-fail}, as decrease in budget is of linear effect, whereas the decreasing of $F$ has an exponential effect as $p_{mc}$ increases. Increasing the fan-out to large numbers but keeping the number of requests needed to fail for the coflow to fail ($F$) the same will result in merely giving additional budget for the attacker.

\paragraphbb{What would a practical coflow attack need?} Our model of coflow-based applications is not entirely realistic. First, it assumes there is only coflow related traffic. There can also be non-coflow background traffic, but this can only help an attack focused on coflows, by allowing the use of loss budget acquired from other traffic on the coflow attack, where this budget is most useful.
Second, the attacker is most effective when in control of a device through which most or all requests for a coflow transit; in particular, with hedging, the attacker needs to be able to see both the request and its replica. This is plausible, for example, if the attacker gains control of the top-of-rack switch under which a user-facing server resides that issues the many backend requests. Third, the attacker needs to achieve reasonable accuracy in classifying packets. Past work~\cite{coflow-coda} has shown that heuristics using only the source and destination addresses and port numbers, and the flow start time, can be effectively used to map flows to coflows with \textbf{over 90\% accuracy}. Identifying hedged requests is also plausible if packets are unencrypted, allowing at least exact matches sent within a certain small time threshold of each other to be easily identified, \eg with the use of bloom filters in P4 switches. The attacker could compare both entire packets, as well as only the payloads, in case the replicated queries are sent to different servers, and thus have different headers. In the presence of encryption, the attacker would need to rely on timing to identify hedged requests; this could also work because a common approach is to simply retry after a short, fixed timeout. 

\section{Mitigation}
\label{sec:mitigation}

Prevention is certainly better than cure, and to that end, standard best practices for network security are certainly helpful. These include safeguarding against physical intrusion~\cite{physical-dc-intrusion}, using responsible distribution of access and authorization, replacement of default device credentials with strong, private credentials\footnote{Sometimes, unfortunately, even industry leading vendors leave no possibility for this~\cite{cisco-advisory-CVE-2016-1329}.}, keeping device software up to date, using a network intrusion detection system~\cite{nids} etc.
However, as is known from past experience, from time to time, such measures will fail, especially in the face of determined adversaries. At least some operators are thus considering the threat of compromised devices even in facilities secured with state-of-the-art security measures (\eg Microsoft, as noted in \S\ref{sec:whyworry}). Thus, a defense-in-depth approach is warranted. Based on our analysis of what enables an \attackName attacker to maximize damage, we can also frame recommendations for mitigating such attacks.

\subsection{Impede classification of key packets}
\label{sec:impede-selectivity}

The success of \attackName attacks is reliant on two factors: (a) the existence of packets which are disproportionately valuable; and (b) the ability of new hardware to identify and operate on such packets selectively.
If either of these two factors is absent, the impact of a \attackName attack reduces drastically. Unfortunately, to some extent, both are fundamentally hard to entirely remove -- the first stems from application and transport structures that are hard to change; and the second is also key to extracting the benefits of such devices, such as in-network telemetry for debugging. We discuss in the following, countermeasures that are increasingly limiting for both the attacker and (as an unfortunate side-effect) the operator:

\begin{itemize}
    \item \vspace{-0.05in}\textbf{Encrypt the payload:} Access to the payload would restrict the application-layer knowledge available to an attacker. For instance, directly identifying redundant or hedged queries by matching payload hashes (\S\ref{sec:application-aware-attacks}) is no longer possible. The attacker would have to resort to likely more inaccurate inferences based on meta-characteristics (\ie timing, size), and as we have seen, limiting the ability of an attacker to classify packets in this way, sharply reduces the attack's effectiveness.
    This method also imposes minimal additional burden on the operator, at least in the data center or enterprise context.
    
    \item \textbf{Encrypt the transport header:} By encrypting parts of the transport header, selecting packets based on \textit{seqno} and \textit{SYN} flags is no longer possible (\S\ref{sec:transport-layer-network-layer-attacks}). Not having access to the source and destination port numbers can also limit identification of coflows (\S\ref{sec:application-aware-attacks}). However, such changes also limit the operator's view; if the attacker cannot identify retransmissions, the operator may no longer be able to do so either, and thus not be able to characterize losses.

    \item \textbf{Obfuscate timing and size:} Both the workload (\eg ``partition aggregate'' of \S\ref{sec:application-aware-attacks}) and the transport protocol state machine can result in predictable sequences and sizes of packets, both in-flow and across flows. By randomly varying timing (\eg as proposed by \cite{lowrateshrew} for $RTO_{min}$) and/or padding of packets, this predictability can be reduced. However, this could potentially increase latency for services.
    
\end{itemize}

\subsection{Application-aware network monitoring}
\label{sec:monitoring-limits}

As we have seen, in their default configuration, existing systems would not be able to detect \attackName attacks. Pushing these systems for greater visibility, \eg by reducing the monitoring interval or increasing sensitivity would in fact limit an attacker, but opens up the risk of more benign false positives as well. Ultimately, the network is a noisy environment, with benign transient congestion, often due to microbursts~\cite{microbursts}, and this noise limits how carefully tuned malicious behavior can be distinguished from this noise. 

Current monitoring systems also struggle with path discovery (which lets such systems ultimately pinpoint which links are faulty) in many approaches~\cite{007, passive-realtime, everflow} in the face of an \attackName attacker who corrupts path-discovery information carried in packets.

These deficiencies are not because these systems are poorly designed -- to the contrary, these are the highest resolution approaches known, in many cases backed by theoretical analysis, with substantial testing and deployment effort behind them. Rather, they are focused on failures, and ill suited for detecting deliberately malicious activity. We believe a promising approach to address the latter problem is to monitor directly the attacker's target: applications.

The \attackName attacker's goal is to cause large and obvious deterioration in the performance networked applications see. Unfortunately, this does not mean that bringing these problems to light is easy. In most cases, the infrastructure will be run by a provider (ISP or cloud data center) for users and application providers. Today, the latter have only slow, manual means for reporting problems they experience. 

At least in some settings, it would be possible to develop an API that lets applications automatically report \emph{their} experience of network services to the operator. For instance, an application can raise an alarm to a cloud operator if its flow or coflow completion time changes substantially. This approach can be viewed as an application of the \textbf{end-to-end argument}~\cite{endendargument} -- only the application or tenant knows whether the network is working well. 

There are also opportunities for the operator to fix the problem in a blackbox fashion: does tunneling this client's traffic (which could make an attacker's targeting harder, or with ECMP, automatically change paths) improve the metrics reported by the client? Such an approach can, incidentally, also work against gray failures. Exploring this possibility, including its potential benefit and its challenges in terms of implementation and stability, is left to future work.

For ISPs, the problem is easier in some dimensions, and much harder in others. It's easier in the sense that the vulnerable coflow structure is largely absent, or manifests with much smaller and less easily detectable coflows (\eg different objects fetched from different servers to compose a Web page). But it's harder in the sense that excepting stub ISPs, an ISP typically has no direct interaction with most end users. It is, unfortunately, unclear to us what countermeasures may be deployable in this setting beyond attempting to track the distribution of congestion window sizes instead of merely coarse bandwidth statistics across aggregates (which is somewhat harder here than in the data center, where monitors benefit from host capabilities), to at least limit the fraction of traffic an attacker can deteriorate.

\section{Related work}
\label{sec:related-work}

We have already discussed recent work on detecting partial failures in \S\ref{sec:monSys}. We thus focus the following discussion on literature related to network attacks and their defenses.

\paragraphbb{Malicious routers:} Prior work (\eg \cite{mizrak-malicious-routers, mizrak-packet-loss}) has analyzed scenarios in which a non-partitioning subset of routers is malicious. This work suggest that each device self report fingerprints of its input-output behavior to the other network devices. These fingerprints are validated to detect inconsistencies. However, this work does not reason about inherent uncertainty in the network (\eg due to congestion or benign failures). It is also unclear whether the model proposed could be translated to practice for today's high port-count, high line-rate devices.

\paragraphbb{Network anomaly detection:} While there is a large body of work on anomaly detection~\cite{nids}, this work largely focuses on the attacks like denial of service, network scans, the spread of viruses and worms, etc., rather than on detecting the performance degradation of legitimate traffic by adversaries who attempt to make their impact as indistinguishable from congestion as possible.

\paragraphbb{TCP attacks:} Several attacks on TCP are well known, such as denial-of-service by targeting TCP retransmissions~\cite{lowrateshrew}, SYN floods that exhaust TCP connect slots~\cite{rfc4987synflood}, and malicious receivers forcing altering the sender's behavior~\cite{misbehavingtcpreceiver}. Our work differs from these attacks in: (a) its goal of making diagnosis hard by budgeting the attack such that it avoids detection by state-of-the-art monitoring systems; (b) its use of in-network programmable devices; and (c) amplifying the effects using application-layer structure.

\paragraphbb{Detecting on-path network misbehavior:} There is also substantial work on detecting which entity on a network path is misbehaving (\eg Network Confessional~\cite{confessional}). However, work in this direction also does not consider a careful attacker who drops or modifies packets at a low enough rate to resemble normal congestive losses. Further, such an attacker, by modifying packet TTLs, can actually blame other on-path entities for packet drops.

\section{Conclusion \& Future work}
\label{sec:discussion-future-work}

We present a new breed of network attack, an \attackName attack, that attempts to cause the maximum performance degradation possible while remaining hard to diagnose. Our proposed attacks are easily implementable in programmable hardware that's on the horizon, as we show through our implementation in P4. We show that \attackName attackers must tread a delicate balance between causing service degradation and getting their compromised devices detected. We also describe how current transport and application primitives, combined with careful budgeting by such an attacker, make it easy to cause disproportionate damage to networked applications without allowing state-of-the-art monitoring systems to identify the compromised devices. 

Our investigation represents the first steps in this new direction, opening up several avenues for future work. We expect that future work will uncover increasingly creative attacks, with the potential to cast blame on benign devices and triggering even more harmful actions from monitoring and response systems. In turn, we expect that new methods of monitoring and securing the network will also be developed to combat such threats.

{\footnotesize \bibliographystyle{acm}
\bibliography{paper.bib}}

\begin{thebibliography}{10}

\bibitem{fattree}
{\sc Al-Fares, M., Loukissas, A., and Vahdat, A.}
\newblock A scalable, commodity data center network architecture.
\newblock {\em SIGCOMM Comput. Commun. Rev. 38}, 4 (Aug. 2008), 63–74.

\bibitem{pfabric}
{\sc Alizadeh, M., Yang, S., Sharif, M., Katti, S., McKeown, N., Prabhakar, B.,
  and Shenker, S.}
\newblock pfabric: Minimal near-optimal datacenter transport.
\newblock {\em ACM SIGCOMM CCR 43}, 4 (2013).

\bibitem{confessional}
{\sc Argyraki, K., Maniatis, P., and Singla, A.}
\newblock Verifiable network-performance measurements.
\newblock {\em ACM CoNEXT\/} (2010).

\bibitem{007}
{\sc Arzani, B., Ciraci, S., Chamon, L., Zhu, Y., Liu, H., Padhye, J., Loo,
  B.~T., and Outhred, G.}
\newblock 007: Democratically finding the cause of packet drops.
\newblock {\em USENIX NSDI\/} (2018).

\bibitem{bmv2}
Behavioral model (bmv2).
\newblock \url{https://github.com/p4lang/behavioral-model}.

\bibitem{nids}
{\sc Bhuyan, M.~H., Bhattacharyya, D.~K., and Kalita, J.~K.}
\newblock Network anomaly detection: methods, systems and tools.
\newblock {\em IEEE Communications Surveys \& Tutorials 16}, 1 (2014).

\bibitem{coflow}
{\sc Chowdhury, M., and Stoica, I.}
\newblock Coflow: A networking abstraction for cluster applications.
\newblock {\em ACM HotNets\/} (2012).

\bibitem{cisco-advisory-CVE-2016-1329}
Cisco nexus 3000 series and 3500 platform switches insecure default credentials
  vulnerability.
\newblock
  \url{https://tools.cisco.com/security/center/content/CiscoSecurityAdvisory/cisco-sa-20160302-n3k},
  March 2016.
\newblock Accessed: 8 September 2018.

\bibitem{cisco-advisory-CVE-2017-6639}
Cisco prime data center network manager debug remote code execution
  vulnerability.
\newblock
  \url{https://tools.cisco.com/security/center/content/CiscoSecurityAdvisory/cisco-sa-20170607-dcnm1},
  June 2017.
\newblock Accessed: 13 September 2018.

\bibitem{cisco-advisories}
Cisco security advisories and alerts.
\newblock
  \url{https://tools.cisco.com/security/center/publicationListing.x?product=Cisco}.
\newblock Accessed: 8 September 2018.

\bibitem{netflow}
{\sc Claise, B.}
\newblock Cisco systems netflow services export version 9, 2004.

\bibitem{tail-at-scale}
{\sc Dean, J., and Barroso, L.~A.}
\newblock The tail at scale.
\newblock {\em ACM SIGCOMM CCR 56}, 2 (2013).

\bibitem{rfc4987synflood}
{\sc Eddy, W.}
\newblock Tcp syn flooding attacks and common mitigations (rfc 4987).
\newblock Tech. rep., IETF, August 2007.

\bibitem{nsa-device-intercept}
{\sc Gallagher, S.}
\newblock Photos of an nsa “upgrade” factory show cisco router getting
  implant.
\newblock
  \url{https://arstechnica.com/tech-policy/2014/05/photos-of-an-nsa-upgrade-factory-show-cisco-router-getting-implant/},
  May 2014.
\newblock Accessed: 8 September 2018.

\bibitem{dapper}
{\sc Ghasemi, M., Benson, T., and Rexford, J.}
\newblock Dapper: Data plane performance diagnosis of tcp.
\newblock {\em ACM SOSR\/} (2017).

\bibitem{pingmesh}
{\sc Guo, C., Yuan, L., Xiang, D., Dang, Y., Huang, R., Maltz, D., Liu, Z.,
  Wang, V., Pang, B., Chen, H., et~al.}
\newblock Pingmesh: A large-scale system for data center network latency
  measurement and analysis.
\newblock {\em ACM SIGCOMM CCR 45}, 4 (2015).

\bibitem{gray-achilles-heel}
{\sc Huang, P., Guo, C., Zhou, L., Lorch, J.~R., Dang, Y., Chintalapati, M.,
  and Yao, R.}
\newblock Gray failure: The achilles' heel of cloud-scale systems.
\newblock {\em ACM HotOS\/} (2017).

\bibitem{thousandCoflows}
{\sc Jalaparti, V.}
\newblock {\em Improving the end-to-end latency of datacenter applications
  using coordination across application components}.
\newblock PhD thesis, University of Illinois at Urbana-Champaign, 2015.

\bibitem{physical-dc-intrusion}
{\sc Jones, R.}
\newblock Five ways to (physically) hack a data center.
\newblock
  \url{https://www.darkreading.com/five-ways-to-(physically)-hack-a-data-center/d/d-id/1133615?},
  May 2010.
\newblock Presented at YSTS. Accessed: 8 September 2018 (news article by Kelly
  Jackson Higgins on 17 May 2010).

\bibitem{mikrotik-used}
{\sc Kasiviswanathan, K.~C.}
\newblock Postmortem of a compromised mikrotik router.
\newblock
  \url{https://www.symantec.com/blogs/threat-intelligence/hacked-mikrotik-router},
  August 2018.
\newblock Accessed: 13 September 2018.

\bibitem{kheirkhah2016mmptcp}
{\sc Kheirkhah, M., Wakeman, I., and Parisis, G.}
\newblock {MMPTCP: A multipath transport protocol for data centers}.
\newblock In {\em IEEE INFOCOM 2016 - The 35th Annual IEEE International
  Conference on Computer Communications\/} (April 2016), pp.~1--9.

\bibitem{russia-hacking}
{\sc Kobie, N.}
\newblock Nobody is safe from russia's colossal hacking operation.
\newblock
  \url{https://www.wired.co.uk/article/russia-hacking-russian-hackers-routers-ncsc-uk-us-2018-syria},
  April 2018.
\newblock Accessed: 12 September 2018.

\bibitem{lowrateshrew}
{\sc Kuzmanovic, A., and Knightly, E.~W.}
\newblock Low-rate tcp-targeted denial of service attacks: the shrew vs. the
  mice and elephants.
\newblock {\em ACM SIGCOMM\/} (2003).

\bibitem{luckymouse}
{\sc Legezo, D.}
\newblock Luckymouse hits national data center to organize country-level
  waterholing campaign.
\newblock
  \url{https://securelist.com/luckymouse-hits-national-data-center/86083/},
  June 2018.
\newblock Kaspersky lab. Accessed: 12 September 2018.

\bibitem{flowradar}
{\sc Li, Y., Miao, R., Kim, C., and Yu, M.}
\newblock Flowradar: A better netflow for data centers.
\newblock {\em USENIX NSDI\/} (2016).

\bibitem{silkroad}
{\sc Miao, R., Zeng, H., Kim, C., Lee, J., and Yu, M.}
\newblock Silkroad: Making stateful layer-4 load balancing fast and cheap using
  switching {ASICs}.
\newblock In {\em {ACM SIGCOMM}\/} (2017).

\bibitem{mizrak-malicious-routers}
{\sc Mizrak, A.~T., Cheng, Y.-C., Marzullo, K., and Savage, S.}
\newblock Detecting and isolating malicious routers.
\newblock {\em IEEE TDSC 3}, 3 (2006).

\bibitem{mizrak-packet-loss}
{\sc Mizrak, A.~T., Savage, S., and Marzullo, K.}
\newblock Detecting malicious packet losses.
\newblock {\em IEEE TPDS 20}, 2 (2009).

\bibitem{hotnets-dialogue-malicious}
{\sc Mogul, J., and Padhye, J.}
\newblock {HotNets-XVI Dialogue:} in-network computation is a dumb idea whose
  time has come.
\newblock
  \url{https://conferences.sigcomm.org/hotnets/2017/dialogues/dialogue140.pdf},
  2017.

\bibitem{cisco200khacked}
{\sc Nandikotkur, G.}
\newblock 200,000 cisco network switches reportedly hacked.
\newblock
  \url{https://www.bankinfosecurity.com/200000-cisco-network-switches-reportedly-hacked-a-10788},
  April 2018.
\newblock Accessed: 12 September 2018.

\bibitem{facebook-memcache}
{\sc Nishtala, R., Fugal, H., Grimm, S., Kwiatkowski, M., Lee, H., Li, H.~C.,
  McElroy, R., Paleczny, M., Peek, D., Saab, P., et~al.}
\newblock Scaling memcache at facebook.
\newblock {\em USENIX NSDI\/} (2013).

\bibitem{ns3}
{\sc ns~3~contributors}.
\newblock ns-3: a discrete-event network simulator for internet systems.
\newblock \url{https://www.nsnam.org/}, 2020.

\bibitem{sflow}
{\sc Panchen, S., Phaal, P., and McKee, N.}
\newblock Inmon corporation's sflow: A method for monitoring traffic in
  switched and routed networks, 2001.

\bibitem{sflowsamplingrates}
{\sc Phaal, P.}
\newblock sflow sampling rates.
\newblock \url{https://blog.sflow.com/2009/06/sampling-rates.html}, June 2009.
\newblock Accessed: 10 September 2018.

\bibitem{defcon-whitebox}
{\sc Pickett, G.}
\newblock Staying persistent in software defined networks.
\newblock {\em DefCon\/} (2015).

\bibitem{passive-realtime}
{\sc Roy, A., Zeng, H., Bagga, J., and Snoeren, A.~C.}
\newblock Passive realtime datacenter fault detection and localization.
\newblock {\em USENIX NSDI\/} (2017).

\bibitem{endendargument}
{\sc Saltzer, J.~H., Reed, D.~P., and Clark, D.~D.}
\newblock End-to-end arguments in system design.
\newblock {\em ACM TOCS 2}, 4 (1984).

\bibitem{misbehavingtcpreceiver}
{\sc Savage, S., Cardwell, N., Wetherall, D., and Anderson, T.}
\newblock Tcp congestion control with a misbehaving receiver.
\newblock {\em ACM SIGCOMM CCR 29}, 5 (1999).

\bibitem{jupiter-rising}
{\sc Singh, A., Ong, J., Agarwal, A., Anderson, G., Armistead, A., Bannon, R.,
  Boving, S., Desai, G., Felderman, B., Germano, P., Kanagala, A., Provost, J.,
  Simmons, J., Tanda, E., Wanderer, J., Hölzle, U., Stuart, S., and Vahdat,
  A.}
\newblock Jupiter rising: A decade of clos topologies and centralized control
  in google’s datacenter network.
\newblock {\em ACM SIGCOMM\/} (2015).

\bibitem{turboflow}
{\sc Sonchack, J., Aviv, A.~J., Keller, E., and Smith, J.~M.}
\newblock Turboflow: information rich flow record generation on commodity
  switches.
\newblock {\em ACM EuroSys\/} (2018).

\bibitem{vera}
{\sc Stoenescu, R., Dumitrescu, D., Popovici, M., Negreanu, L., and Raiciu, C.}
\newblock Debugging p4 programs with vera.
\newblock {\em ACM SIGCOMM\/} (2018).

\bibitem{netbouncer}
{\sc Tan, C., Jin, Z., Guo, C., Zhang, T., Wu, H., Deng, K., Bi, D., and Xiang,
  D.}
\newblock Netbouncer: Active device and link failure localization in data
  center networks.
\newblock In {\em 16th {USENIX} Symposium on Networked Systems Design and
  Implementation ({NSDI} 19)\/} (Boston, MA, Feb. 2019), {USENIX} Association,
  pp.~599--614.

\bibitem{nsa-genie}
{\sc Zetter, K.}
\newblock {NSA} laughs at {PC}s, prefers hacking routers and switches.
\newblock \url{https://www.wired.com/2013/09/nsa-router-hacking/}, September
  2013.
\newblock Accessed: 12 September 2018.

\bibitem{mikrotik-hacked}
{\sc Zetter, K.}
\newblock Unpatched routers being used to build vast proxy army, spy on
  networks.
\newblock
  \url{https://arstechnica.com/information-technology/2018/09/unpatched-routers-being-used-to-build-vast-proxy-army-spy-on-networks/},
  September 2018.
\newblock Accessed: 13 September 2018.

\bibitem{coflow-coda}
{\sc Zhang, H., Chen, L., Yi, B., Chen, K., Chowdhury, M., and Geng, Y.}
\newblock Coda: Toward automatically identifying and scheduling coflows in the
  dark.
\newblock {\em ACM SIGCOMM\/} (2016).

\bibitem{microbursts}
{\sc Zhang, Q., Liu, V., Zeng, H., and Krishnamurthy, A.}
\newblock High-resolution measurement of data center microbursts.
\newblock {\em ACM IMC\/} (2017).

\bibitem{everflow}
{\sc Zhu, Y., Kang, N., Cao, J., Greenberg, A., Lu, G., Mahajan, R., Maltz, D.,
  Yuan, L., Zhang, M., Zhao, B.~Y., et~al.}
\newblock Packet-level telemetry in large datacenter networks.
\newblock {\em ACM SIGCOMM CCR 45}, 4 (2015).

\end{thebibliography}

\end{document}